\documentclass[a4paper,11pt]{article}
\usepackage{pos}
\usepackage[capitalise]{cleveref}

\title{Towards the new NNPDF4.1 global analysis}

\author*[a,b]{Felix Hekhorn}

\affiliation[a]{University of Jyvaskyla, Department of Physics, P.O. Box 35, FI-40014 University of Jyvaskyla, Finland}
\affiliation[b]{Helsinki Institute of Physics, P.O. Box 64, FI-00014 University of Helsinki, Finland}

\emailAdd{felix.a.hekhorn@jyu.fi}

\abstract{We review progress and plans towards a new global PDF analysis from the NNPDF Collaboration.
This new release will be based on state-of-the-art theory calculations, and in particular will be based on exact NNLO interpolation grids, (approximate) N3LO QCD corrections to DGLAP evolution and DIS structure functions, and QED effects to DGLAP evolution.
All relevant sources of experimental and theoretical uncertainties, the latter including missing higher orders and higher twists, are taken into account.
As compared to NNPDF4.0, this new release includes more than 25 new datasets from HERA jets to Drell-Yan (inclusive and with extra jets), top quark, and jet production at the LHC among others.
Our machine learning methodology is statistically validated by extensive closure and future tests and benefits from an improved hyperoptimization algorithm.}

\FullConference{The 33rd International Workshop on Deep Inelastic Scattering and Related Subjects (DIS2026)\\
4 - 8 May 2026\\
Bologna, Italy\\}

\begin{document}
\maketitle

\section{Introduction}
The last major PDF release of the NNPDF Collaboration, NNPDF4.0~\cite{NNPDF:2021njg}, dates now five years back and in the meantime we have made a number of significant advances on all three pillars of PDF extraction: theory setup, experimental data, and methodological tools.
The full combination of all improvements will culminate in an upcoming new PDF release, dubbed NNPDF4.1.
We review in this contribution a selected set of enhancements, which have been studied by the collaboration or a smaller subgroup thereof.
We focus on the specific advancement for global PDF extractions and highlight the different directions we are exploring.

Before we discuss the advances in the three main pillars in turn, it is worth recalling again that all developments inside the NNPDF Collaboration are available as open source repositories~\cite{NNPDF:2021uiq}\footnote{see \url{https://github.com/NNPDF}}.
In particular, this includes software programs to compute theory predictions (e.g.\ \texttt{PineAPPL} or \texttt{EKO}), to access experimental measurements (e.g.\ \texttt{nnpdf\_data}), or to perform the actual fitting (e.g. \texttt{validphys2} or \texttt{n3fit}).
With the ever increasing complexity, for which we give explicit examples below, the availability of open source codes becomes increasingly critical and, up to this day, only the NNPDF Collaboration has taken this important step.

\section{Theory advances}
The computation of theoretical predictions has been completely revised since our last major release, culminating in the introduction of the \texttt{pineline} framework~\cite{Barontini:2023vmr}, which aims at an efficient and reproducible calculation of theory predictions.
In particular, we need to consider the evolution of PDF and the computation of partonic matrix elements, which we can combine eventually together into a fast-kernel (FK) table~\cite{Barontini:2023vmr}.

We have developed the open-source \texttt{EKO} library~\cite{Candido:2022tld} for PDF evolution, which has been extended from its initial implementation in a range of papers.
Thanks to the possibility of evolving all PDFs to any scale, we were able to demonstrate first evidence for the existence of an intrinsic charm component in the proton~\cite{Ball:2022qks} and we could then show that a measurement of an charm asymmetry would yield further tight constraints~\cite{NNPDF:2023tyk}.
In Ref.~\cite{NNPDF:2024djq} we added the possibility to allow the photon to participate in PDF evolution and, moreover, we realized that in order to so faithfully we need to adopt the so-called exact solution~\cite{Bonvini:2012sh}, which is often also used by other PDF fitting groups.
After having revised the strategy to include Missing Higher Order Uncertainties (MHOUs) in PDF extractions~\cite{NNPDF:2024dpb}, we could then extend our perturbative accuracy up to next-to-next-to-next-to-leading order (N$^3$LO) accuracy in the strong coupling~\cite{NNPDF:2024nan}.

The computation of partonic matrix elements evolves centrally around our fast-interpolation grid library PineAPPL~\cite{Jezo:2026adf}.
The open-source \texttt{Yadism} library~\cite{Candido:2024rkr} provides coefficient functions for DIS up to N$^3$LO including the recent progress on heavy quark production~\cite{Barontini:2026drp}.
Instead, thanks to the interface to both Matrix~\cite{Devoto:2025cuf} and NNLOJET~\cite{NNLOJET:2025rno} we obtain, for the first time, access to partonic matrix elements in hadronic collisions, which are exact up to NNLO accuarcy and which may help in flavor separation~\cite{Cruz-Martinez:2025ffa}.

While reviewing our entire framework for theory predictions we also address further issues.
We revise our choice of the central scale in several hadronic measurement, updating it to the more recent recommendations.
We improve our treatment of electro-weak (EW) corrections for all included measurements by considering a wide range of EW K-factors and adopting a consistent EW input parameter scheme, the $G_\mu$ scheme~\cite{Denner:2019vbn}.

\section{New experimental data}
\begin{figure}[ht]
    \centering
    \includegraphics[width=.7\textwidth]{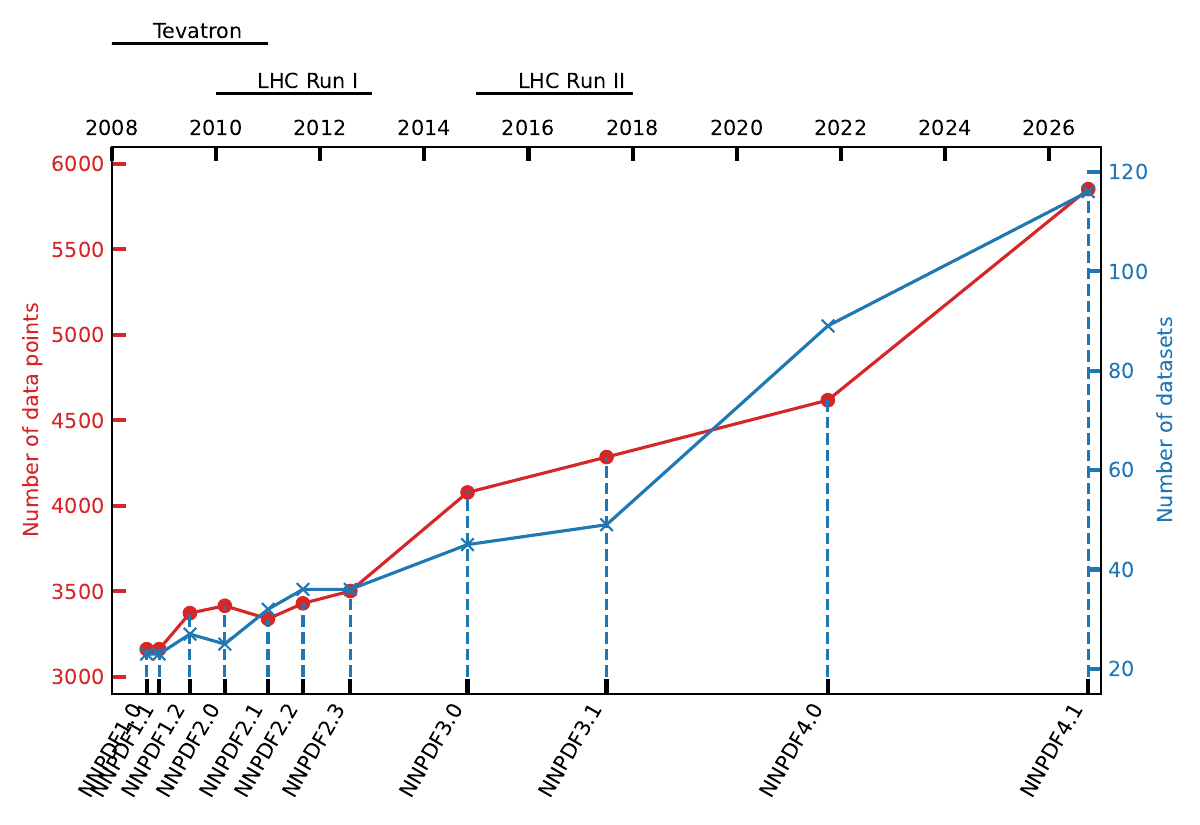}
    \caption{The evolution in number of data points (red lines, left y-axis) and in number of datasets (blue lines, right y-axis) in subsequent NNPDF releases from NNPDF1.0 to NNPDF4.1, spaced by the corresponding year of release.
    The number for NNPDF4.1 is merely indicative as an excessive upper limit including all potentially available data points, which, however, are still subject to double-counting and assessment to fit quality.}
    \label{fig:data}
\end{figure}
We have completely revised our treatment of experimental data since our last major release and, while doing so, have fixed a number of minor inconsistencies.
In addition, we extend our dataset by a vast number of diverse measurements, summarized in \cref{fig:data}.
Note that the indicated numbers for NNPDF4.1 are merely indicative as an excessive upper limit including all potentially available data points, which, however, are still subject to double-counting and assessment to fit quality.
We impose as a cut-off rule, that eligible datasets must have appeared before January 1st, 2025 on the HEPData repository~\cite{Maguire:2017ypu}.

A significant fraction of the new experimental data was assessed in Ref.~\cite{Chiefa:2025loi} finding consistency between most modern PDF extractions.
The scrutinized data includes Drell-Yan and top quark measurements at the LHC as well as single jet and dijet measurements from both, the LHC and HERA.
The top quark measurements were further investigated in Ref.~\cite{Ball:2026qno}, which demonstrated how the availability of several mutually dependent measurements of a single experimental event can and will complicate the consolidation of the finally adopted dataset.
The role of higher twist and power corrections was investigated in Ref.~\cite{Ball:2025xtj} and, on the one side, confirmed the existing kinematic cuts to DIS and jet measurements, and, on the other hand, showed how to account for the associated uncertainty in a PDF fit.

Finally, in Ref.~\cite{Barontini:2025lnl} we address the faithful construction of a dataset suitable for PDF extraction more directly.
Using the closure test framework, we find that the neural network (NN) approach can compensate for mild data inconsistencies and propose an algorithm for cases where this fails.
In particular, we check for the impact of a given dataset both in the fit itself as well as on the description of any other dataset to maximize consistency.

\section{Methodological advances}
The successful adaption of our core fitting algorithm onto Graphical Process Units (GPUs) demonstrated in Ref.~\cite{Cruz-Martinez:2024wiu} yields a significant speedup and allows us to further improve the fitting procedure.
In particular, Ref.~\cite{Cruz-Martinez:2026aqe} suggested an improved hyperoptimization strategy, which, effectively, adds an additional uncertainty associated to the architecture of the fitting NN itself by sampling over statistically equivalent configurations.
However, we stress that all generated replicas are still subject to the full fitting algorithm, which includes, e.g., dedicated steps to prevent overfitting, such as a training-validation separation.

Following Ref.~\cite{NNPDF:2023tyk} we will parametrize the charm and anti-charm quark distributions independently, as for any other light flavor, which then yields nine independent PDFs at the starting scale.
The positivity of $\overline{\mathrm{MS}}$ PDFs was proven in Ref.~\cite{Candido:2023ujx} given perturbative QCD can be applied.
In order to enforce this condition stronger, we will revise our positivity constraints and impose them henceforth in the large $Q^2$ and large $x$ region.
The large $x$ behavior was further investigated in Ref.~\cite{Carrazza:2021yrg}, where we remove the pre-pocessing polynomial, i.e., schematically, we can write our PDFs at parametrization scale $Q_0$ as
\begin{equation}
    f^\text{NNPDF4.1}_j(x,Q_0^2) = x^{\alpha_j} \cdot \mathrm{NN}_j(x)
\end{equation}
and the large $x$ behavior is captured by a suitable recalling of the NN.

\section{Conclusion}
Since the release of NNPDF4.0~\cite{NNPDF:2021njg}, the NNPDF Collaboration has revised and improved all three major pillars of an PDF extraction: theory predictions, experimental data, and methodological tools.
Many features of this diverse list have been presented in various dedicated papers, which will, finally, combined into the new upcoming PDF release.
We will employ the currently best available theory with exact NNLO partonic matrix elements or better if possible.
The PDF evolution will be performed by default at N$^3$LO QCD $\otimes$ NLO QED accuarcy~\cite{Barontini:2024eii}.
We consider new experimental measurements coming from either LHC or HERA up to January 1st, 2025, while using any later measurements to validate our PDFs, for example in so-called future tests~\cite{Cruz-Martinez:2021rgy}.
The potentially available data points amount up to 1500 measurements coming from up to 25 different datasets.
We apply the algorithms of Refs.~\cite{Ball:2025xtj,Cruz-Martinez:2026htl} to select a range of valid data points, which are admissible for a faithful PDF extraction.
We start to parametrize both, charm and anti-charm, independently and remove the large $x$ pre-processing for all quarks.
We diversify the architecture of the fitting NN by using a more advanced hyperoptimization algorithm.
Of course, the closure test framework will remain an integral part of our methodological tools and its importance was highlighted in the extraction of the strong coupling in Ref.~\cite{Ball:2025xgq}.

\bibliographystyle{utphys}
\bibliography{refs.bib}

\end{document}